      [1999/12/01 v1.4c Il Nuovo Cimento]
\documentclass{cimento}

\usepackage{graphicx}  % got figures? uncomment this
\title{SN/GRB connection: a statistical approach with BATSE and Asiago 
Catalogues}

\author{S.~Valenti\from{ins:x}\from{ins:y}
E.~Cappellaro\from{ins:y},
M.~Della Valle\from{ins:z},
F.~Frontera\from{ins:x}\from{ins:w}\\
C.~Guidorzi\from{ins:zz} 
        \atque
E.~Montanari\from{ins:x}
}
\instlist{
\inst{ins:x} Physics Department, University of Ferrara, I-44100 Ferrara, Italy
\inst{ins:y} INAF - Astronomical Observatory of Capodimonte, Via Moiariello 16,I-80131 Napoli,Italy
\inst{ins:z} INAF - Astronomical Observatory of Arcetri, Via Enrico Fermi 5, I-50125 Florence, Italy
\inst{ins:w} INFN,IASF, Via Gobetti, 101, 40129 Bologna, Italy
\inst{ins:zz} Astrophysics Research Institute - Liverpool
  JM University, Twelve Quays House, Egerton Wharf, Birkenhead CH41 1LD, United Kingdom
}
\PACSes{\PACSit{88.70} GAMMA-RAY BURST}
\begin{document}

\maketitle

\begin{abstract}
Recent observations suggest that some types of GRB are physically
connected with SNe of type Ib/c. However, it has been pointed out by
several authors that some GRBs could be associated also with other
types of core-collapse SNe (type IIdw/IIn). On the basis of a
comphrensive statistical study, which has made use of the BATSE and
Asiago catalogues, we have found that: {\sl i)} the temporal and spacial
distribution of SNe-Ib/c is marginally correlated with that of the
BATSE GRBs; ii) we do not confirm the existence of an association
between GRBs and SNe-IIdw/IIn.
\end{abstract}

\section{Introduction:}

The association between Supernovae of type Ib/c (SNe-Ib/c)
and long duration GRBs has been firmly established only in a few cases
(see Della Valle et al. 2005 for a review). However, there have been
claims, based on spatial and temporal SN-GRB coincidencies, that also some
other types of core-collapse SNe, the so called SNe-IIdw (see Hamuy
2003 for a review) might be associated with GRBs (Germany et al. 2000,
Turatto et al. 2000, Rigon et al. 2003).  To test this idea we have
correlated the GRBs from BATSE catalogue (2702 GRBs from April 1991 to
May 2000) with SNe of the Asiago catalogue detected in the same period
of BATSE (736 SNe).

\section{Correlation between Asiago SNe and BATSE GRBs catalogues:}
\label{intro}

We selected all pairs GRB-SN from the BATSE and Asiago catalogues that
match the following spatial and temporal requirements:
\begin{itemize}

\item  all SNe falling within one sigma error radii of BATSE error-boxes 
increased with the sistematic error of the BATSE instrument of 1.85 degrees
(Briggs et al. 1999). This choice is motivated by the fact that the typical
size of the error box associated with GRBs, varies from 0.5 to several
degrees, which is much larger than the `average' error box ($\sim$ a
few arcsec) associated with SNe of the Asiago archive.

\item we assumed, on the basis of the SN/GRB associations so far discovered, 
that SNe and GRBs go off simultaneously.  Thus the temporal window is
set by the lenght of the rising time of the SN from the epoch of the
explosion to maximum light. This parameter is not well known, however
we can assume, on the basis of SN 1998bw, $T_{rise}\sim 15$ days. 
Given the uncertainty affecting the epoch of the SN maximum 
light, we assume $T_{rise}\pm 11$ days.
\end{itemize}

\section{Preliminary Analysis}
\label{s:data}

In Table {\bf Ia}, we report, for each SN type, the number (N(a)) and 
the fraction (\%) of GRB-SN pairs, 
obtained by searching for each GRB error-box, the SNe possibly exploded
in that error-box, with maximum light at $15^d \pm 11^d$ 
afther GRB occurence.

These figures have to be compared with the analogous entries in Table {\bf Ib},
where we report, for each type of SN, the number (N(b)) and 
the fraction (\%) of SNe included in the Asiago catalogue. 
If SNe and GRBs are not physically connected, the
temporal constraints will not introduce any bias in forming SN/GRB
pairs. In other words, one should expect to find similar or lower frequencies
 of SNe in Table {\bf Ia} and in Table {\bf Ib}.
This is certainly the case for type
Ia, normal II, IIb, IIn, whereas for type Ib/c the pairs frequency is
about three times as much as expected, after assuming the pair
frequency due to random coincidences.  Simple applications of
Poissonian statistic in regime of small numbers (Gehrles 1986) suggest
a significance of the order of $\sim 93 \%$. Note that we did not
include in our analysis SN 1998bw.
\begin{table}[!ht]
\caption{$\bf{(a)}$ 
The number and the frequency for each type of SN, by correlating 
 Asiago and BATSE catalogues (see text) 
In the sample of SNe we ignored the association (SN1998bw-GRB980425);
  $\bf{(b)}$ The number and frequency for each type of SN included in the Asiago catalogue in the temporal window of BATSE (1991-2000).}
  \label{tab:pricesI}
    \begin{tabular}{c|c|c|c|c|c|c}
    \hline
$\bf{SNe}$  & $\bf{Ia}$ & $\bf{II}$ & $\bf{IIb}$ & $\bf{IIn}$ & $\bf{Ibc}$ & $\bf{Tot.}$ \\
\hline
$\bf{N(a)}$  & 15  & 12 & 0 & 3 & 6 & 36 \\
$\bf{\%}$  & 41.7  &  33.3 & 0 & 8.3 & 16.6 & 100 \\
\hline
\hline
$\bf{SNe}$  & $\bf{Ia}$ & $\bf{II}$ & $\bf{IIb}$ & $\bf{IIn}$ & $\bf{Ibc}$ & $\bf{Tot.}$ \\
\hline
$\bf{N(b)}$  & 448  &  202 & 4 & 38 & 44 & 736 \\
$\bf{\%}$ & 60.9  &  27.4 & 0.5 & 5.2 & 6.0 & 100 \\
    \hline
  \end{tabular}
\end{table}

Afterwards we have repeated the cross-correlation by varying the time
windows (i.e. $T_{rise}\pm \Delta T/2$; with $\Delta T = 22,
30$... days) and computed the statistical significance, of the number
of pairs obtained after each run, with a Montecarlo simulation. 
For the result of this simulation, see the Discussion. 

\section{The Montecarlo Simulations}

We built up 1000 Simulated Samples of SNe (SS) by keeping the SN
positions in the sky (according to the SN catalogue) and changing
randomly the SN types. We have repeated this procedure for 14 temporal
windows and we have computed the probability to obtain, by chance, the
distributions of the SN-GRB pairs actually observed. The results are
shown in Fig 1.

\begin {figure}[!h]
\centering
\begin{center}
\includegraphics [ width=5.9cm, height=7.cm, angle=270]{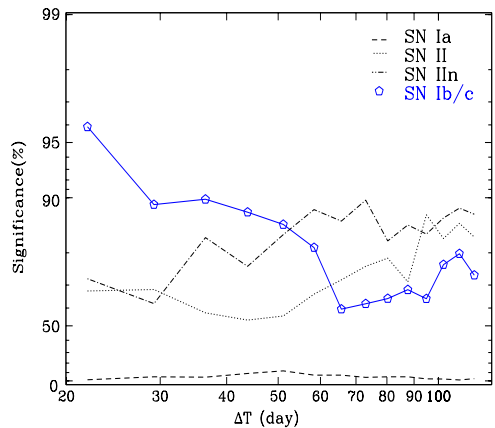}
\caption{Statistical significance of the frequencies of occurence of the
 pairs SN/GRB, for different classes of SNe, as a 
 function of the Temporal Window ($= T_{rise} \pm \Delta T/2$).}  
\label{figura}
\end{center}
\end{figure}

\section{Discussion}
\label{s:discussion}

We can summarize our results as follows:

1. None of the trends shown in Fig. 1 is statistically significant
   with the possible exception of SNe-Ib/c for which, in
   correspondence of a temporal window 26--4 days before maximum
   ($\Delta T = 22$, the first point of the figure), the frequency of
   Ib/c obtained with the Asiago and BATSE archive is significant at
   the level of $96\%$, consistently with our preliminary analysis
   (see sect. 3). Our findings confirm the results of Wang and Wheeler
   (1998) and disagree with the conclusions of Kippen et al. 1998
   (both works have used smaller samples of GRBs and SNe).

2. The other types of SNe do not show any physical connection with
   GRBs. In particular we do not find any evidence for the existence
   of a  physical  association between GRBs and SN-IIdw as proposed on
   the basis of spatial and temporal coincidencies for GRB 970514/SN
   1997cy (Germany et al. 2000, Turatto et al. 2000) and GRB 980910/SN
   1999E (Rigon et al. 2003). Our results suggest that SNe-IIdw
   are not the major class of progenitors for GRBs, although one can
   not exclude that sporadically SNII-dw are able to produce GRBs (see
   for example Garnavich et al. 2003).

3. Since most SNe listed in the Asiago catalogue have been discovered
   within $z < 0.1$ our results apply only to SN/GRB associations
   which occur in the `local' universe.

4. A remarkable result of this statistical approach is to obtain a
   restricted sample of SN/GRB associations (see tab. II) worth being
   further investigated.

\begin{table}[!ht]
  \caption{The 6 pairs SN-GRB, with the SN Ib-c, selected with the described method.}
\small
  \label{tab:pricesII}
   \begin{narrowtabular}{0.cm}{llllllll}
   \hline 
SN  &  Type  &  Epoch  &  redshift& GRB  & Epoch &  ErrorBox & Type \\
  & SN  &  SN  &  SN &  &  GRB & GRB & GRB \\
\hline 
1993R  &  Ic &  1993.41 & 0.0055 &  930524 & 17h44m52s & 18.0 & short\\
1996bx &  Ic  &  1996.88   & 0.062  & 961029 & 06h34m37s  & 3.3  & long\\
1997B  &  Ic  &  1997.03   & 0.01   & 961218 & 19h35m31s  & 12.7 & long\\
1997dq &  Hyp &  1997.83 & 0.0032 & 971013 & 08h43m23s  & 8.8  & long\\
1999dn &  Ib-c &  1999.62 & 0.0094 & 990810 & 07h24m53s  & 4.4  &  short\\
1999ex &  Ib-c  &  1999.85 &  0.0114 & 991021 & 01h58m41s  & 12.3 & unknown\\
\hline
  \end{narrowtabular}
\end{table}

%\subsection{Citations}
%We're almost done, just some citations~\cite{ref:apo}
%and we will be over~\cite{ref:pul,ref:bra}.
%
%\appendix
%\section{}
%Let us go then, you and I\ldots

%\acknowledgments
%This work was produced, supported and perpetrated by S. Frabetti under
%the auspices of the Italian Physical Society.
%Grazie to M. Missiroli for the valuable collaboration.

\end{document}